%% file: elsarticle-template-harv.tex
\documentclass[final,5p,times,twocolumn]{elsarticle}

\usepackage{caption}
\usepackage{subcaption}

\usepackage{wrapfig}

\usepackage{xcolor}

\usepackage[latin1]{inputenc}
\usepackage{fancyhdr}
\usepackage{subcaption}
\usepackage{geometry}
\usepackage{makeidx}
\usepackage{graphicx}
\usepackage{latexsym}
\usepackage{eucal}
\usepackage{setspace}
\usepackage{multirow} 
\usepackage{booktabs} 
\usepackage{amsmath}
\usepackage{graphicx}
\usepackage{comment}
\usepackage[autostyle=false, style=english]{csquotes}
\MakeOuterQuote{"}

\usepackage{array, booktabs, makecell}

\usepackage[hyphens]{url}
\begin{document}

\begin{frontmatter}



\title{
COVID-19 Pandemic Outbreak in the Subcontinent: A data-driven analysis} 

 \author[label1]{Bikash Chandra Singh}
 \address[label1]{Dept. of Information and Communication Technology, Islamic University, Kushtia, Bangladesh }

\author[label2]{Zulfikar Alom}
\address[label2]{Dept. of Computer Science, Asian University for Women, Chittogram, Bangladesh}

\author[label3]{Mohammad Muntasir Rahman}
\address[label3]{Dept. of Computer Science and Engineering, Islamic University, Kushtia, Bangladesh }

\author[label4]{Mrinal Kanti Baowaly}
\author[label2]{Mohammad Abdul Azim}
\address[label4]{Dept. of Computer Science and Engineering, Bangabandhu Sheikh Mujibur Rahman Science and Technology University, Gopalganj, Bangladesh}

\begin{abstract}

Human civilization is experiencing a critical situation that presents itself for a new coronavirus disease 2019 (COVID-19). This virus emerged in late December 2019 in Wuhan city, Hubei, China.  
The grim fact of COVID-19 is, it is highly contagious in nature, therefore, spreads rapidly all over the world and causes severe acute respiratory syndrome coronavirus 2 (SARS-CoV-2). Responding to the severity of COVID-19 research community directs the attention to the analysis of COVID-19, to diminish its antagonistic impact towards society. Numerous studies claim that the subcontinent, i.e., Bangladesh, India, and Pakistan, could remain in the worst affected region by the COVID-19. In order to prevent the spread of COVID-19, it is important to predict the trend of COVID-19 beforehand the planning of effective control strategies. Fundamentally, the idea is to dependably estimate the reproduction number to judge the spread rate of COVID-19 in a particular region.
 Consequently, this paper uses publicly available epidemiological data of Bangladesh, India, and Pakistan to estimate the reproduction numbers. More specifically, we use various models (for example, susceptible infection recovery (SIR), exponential growth (EG), sequential Bayesian (SB), maximum likelihood (ML) and time dependent (TD)) to estimate the reproduction numbers and observe the model fitness in the corresponding data set. Experimental results show that the reproduction numbers produced by these models are greater than 1.2 (approximately) indicates that COVID-19 is gradually spreading in the subcontinent.

\end{abstract}




\end{frontmatter}


\input{Introduction}

\input{Related_work}

\input{Backgroud_SIR_SEIR_Model}

\input{Dataset}

\input{Experiments}

\input{Conclusion}




\bibliographystyle{elsarticle-num}

\bibliography{reference-file}

\end{document}

%% file: Introduction.tex
\section{Introduction}\label{Intro}

The coronavirus disease 2019 (COVID-19) pandemic has spread so rapidly around the globe that the modern world could not have imagined its severity at the beginning of the outbreak that started in Wuhan, China in late December 2019. On 05 January 2020, the world health organization (WHO) first released information on the disease outbreak that a cluster of cases of pneumonia caused by an unknown reason was detected in Wuhan city \cite{who2020pneumonia}. After that, the Chinese authorities identified the disease caused by a novel coronavirus \cite{wu2020new} on 07 January 2020. WHO temporarily termed it as 2019-nCoV \cite{WHO2020nCoV} on 12 January 2020; was later officially named as ``severe acute respiratory syndrome coronavirus 2 (SARS-CoV-2)'' by the international committee on taxonomy of viruses (ICTV) based on genetic analysis \cite{ICTV2020species,WHO2020naming}. 

Initially, WHO declared the outbreak as a public health emergency of international concern (PHEIC) \cite{WHO2020emergency} on 30 January 2020. Eventually by evaluating the severity and alarming levels of the transmission of its high contagiousness, WHO characterized the COVID-19 as a pandemic \cite{WHO2020pandemic} on 11 March 2020. With this declaration COVID-19 becomes the fifth recorded pandemic after the 1918 Spanish flu pandemic \cite{liu2020covid}.

As of 30 June 2020, COVID-19 is still in a pandemic with $10,434,890$ confirmed cases and $508,843$ deaths worldwide \cite{worldometer}, and continues to climb globally. Unlike all other pandemics recorded in history, large amounts of data and news about COVID-19 are rapidly spreading and widely reported, and scholars in various fields have been mobilized to concentrate on analyzing these data and proposing solutions. Since the governments of different countries have responded to the COVID-19 pandemic seriously, it is important that the researchers estimate: (i) the pandemic regionally based on the basic reproduction number, (ii) the arrival of the peak time, and forecast the time course of the epidemic by analyzing the data on the total number of infected cases, (iii) the total number of confirmed cases, (iv) the total number of deaths, and (v) the total number of cases recovered, etc. Many researchers around the globe have estimated the prediction of the COVID-19 spreading and the end of the epidemic in different countries \cite{sulaiman2020dynamical,zhao2020estimating,roda2020difficult,lin2020conceptual, tang2020estimation, yang2020modified, fanelli2020analysis, salgotra2020time, djilali2020coronavirus}. Some of these methods are based on statistical models \cite{roosa2020real,li2020propagation,salgotra2020time,acuna2020modeling} and some other methods use deterministic epidemic models known as susceptible infectious recovery (SIR) with different forms \cite{feng2020benefits,sulaiman2020dynamical,dhanwant2020forecasting, bertozzi2020challenges, de2020sir, qi2020model}.

Recently, some countries in South Asia, especially in \\ Bangladesh, India and Pakistan, cases of COVID-19 are increasing rapidly. In particular, the COVID-19 cases became a new hot spot after countries began to relax locking restrictions.
In this study, we use SIR, EG, SB, ML and TD models to analyze data to determine the reproduction number and predict the epidemic trend of COVID-19 in Bangladesh, India and Pakistan. More particularly,  this article investigates the basic reproduction number $R_0$ and effective reproduction number $R(t)$ using these models. Previous studies show that $R_0$ is applicable when an exponentially increasing epidemic starts in the case of a completely susceptible population \cite{shim2020transmission}. 

\textbf{Contributions:} The contribution of this article is twofold: (i) estimating of $R_0$ for Bangladesh, India and Pakistan, (ii) estimating  $R(t)$ which quantifies the transmission potential over time. As the epidemic progresses, this parameter will track the average number of secondary cases per case over time periods.
After calculating the $R_0$ and $R(t)$s, we compare the spread of COVID-19 in the region of concern. Experimental results show that the $R_0$ and $R(t)$s in all these countries are higher than about 1.2.

The rest of the paper is organized as follows. Section
\ref{related_works} discusses related work, while Section \ref{reproduction_number} outlines the concept of reproduction numbers. Section \ref{Epidemic_models} describes several models that can be used to generate reproduction numbers, whereas Section \ref{data} explains information about the data source. Section \ref{experimental_results} explains the experimental results. Finally, Section \ref{conclustion} summarizes the paper.

%% file: Related_work.tex
\section{Related work}\label{related_works}

A number of models have been proposed and applied to the area of infectious COVID-19. The majority of the mathematical models fall into the categories: (i) standard statistical models applied to the COVID-19 dataset, (ii) deterministic epidemic models such as SIR, SEIR etc. (iii) modified variants of the well-known models attempting to incorporate specific criteria such as incorporating social distancing into SEIR, and (iv) effectiveness of the contact tracing and isolation.
 
In~\cite{benvenuto2020application}, authors choose auto regressive integrated moving average (ARIMA) to predict the spread of COVID-19 based on Johns Hopkins University dataset~\cite{Johns}. Authors intend to avoid potential biases of the evaluating model and select a simple and economical model therefore chosen ARIMA. The ARIMA model consists of an autoregressive (AR) model, moving average (MA) model, and seasonal autoregressive integrated moving average (SARIMA) model~\cite{fattah2018forecasting}. Results demonstrate the autocorrelation function (ACF), and the partial autocorrelation function (PACF) prevalence and incidence of COVID-19 are not influenced by the seasonality. However the claim is arguable as the COVID-19 may have seasonality influences; it is just too early to claim.

Calafiore \textit{et al.}~\cite{calafiore2020modified} presents a modified susceptible-infected-recovered (SIR) model for the contagion analysis of COVID-19 in Italy. Besides the regular SIR parameters the authors incorporate the initial number of susceptible individuals. Most interestingly authors consider a proportionality factor i.e., the ratio of the detected number of COVID positives to the actual number of infected individuals. Two distinct algorithms (i) tuning of the model parameters and (ii) predicting the number of infected, recovered and deaths are presented in this article. 

Kucharski \textit{et al}.~\cite{kucharski2020early} reported how transmission in Wuhan varied over a time period (January - February 2020). The article also attempts to put foresight on how human-to-human transmissions may occur outside Wuhan by travellers. The article reported the median day-to-day reproduction number over the time in Wuhan. The susceptible exposed infected and removed (SEIR) model is used for the evaluations. Besides, the study attempts to incorporate the uncertainty of the detection and observation utilizing Poisson process and binomial observation process on infection prevalence on flights.

Peng \textit{et al}.~\cite{peng2020epidemic} attempted to modify the generalized SEIR model by incorporating quarantine into account and studied the COVID-19 pandemic in 24 provinces in Mainland and 16 counties in Hubei province in China. The authors estimate the latent time, the quarantine time, the reproduction number, the possible ending time and the final total infected cases. 

Wangping \textit{et al.}~\cite{wangping2020extended} proposed an extended susceptible infected removed (eSIR) model essentially an extension to the SIR model attempting to address the effects of different intervention measures in dissimilar periods. The eSIR is applied to model the pandemic COVID-19 analysis in Italy. The markov chain monte carlo (MCMC) algorithm is utilized to obtain the posterior estimation of the unknown parameters in the SIR model.
 
Chatterjee \textit{et al.}~\cite{ chatterjee2020healthcare} modeled a variant of SEIR for COVID-19 epidemic in India. In this stochastic modeling approach Monte Carlo simulation is used to model the concept with a 1000 run. The article shed into the hospitalization and intensive care unit (ICU) requirements along with deaths. Article also presents the impact of the lockdown and social distancing.

Liang \textit{el al.}~\cite{liang2020mathematical} presents infection kinetic analysis of SARS, MARS and COVID-19. Author attempts to acquire a propagation growth model by utilizing the growth rate, and inhibition constant of the aforementioned diseases. The inhibition constant depends on the prevention and control measures adopted for the population. The article assumes during the inception of the disease cycle no effective measure is in place. It figures out the growth rate, the multiplication cycle and infection inhibition constant.

Ndairou \textit{et al.}~\cite{ndairou2020mathematical} provides a mathematical model i.e., essentially an extension to the SIR model of COVID-19 taking eight epidemiological classes into consideration. They are (i) susceptible, (ii) exposed, (iii) symptomatic and infectious, (iv) super spreaders, (v) infectious but asymptomatic, (vi) hospitalized, (vii) recovery, and fatality class. Wuhan number of cases and deaths are compared with the model.

Roda \textit{et al.}~\cite{hellewell2020feasibility} argued that these model predictions have shown a wide range of variations. And the variation of the performances of the models is due to the non-identify ability in model calibrations. The qualities of the statistical models are studied utilizing the Akaike information criterion (AIC). Authors' finding the performance of SIR over SEIR is explained as the failure of the even more sophisticated model is due to the more challenging realization of a comparatively more complex model.
 
Contrary to the other school of analytical models where the models used to predict the number of cases and deaths. Hellewell \textit{et al.}~\cite{hellewell2020feasibility} attempts to quantify the effectiveness of the contact tracing and isolation to control the COVID-19. The study considers a number of scenarios containing (i) initial cases, (ii) $R_0$, (iii) delay in-between symptom and isolation, (iv) probability of traced contacts, (v) proportion of transmission before symptom, and (vi) proportion of sub-clinical infections.

Contrarily this study attempts to find the transmission potential in the subcontinent. Instead of relying on just one specific model this exploration employs SIR, EG, SB, ML and TD that realize definitive regional $R_0$ and $R(t)$ that get a conclusive outcome of the containment measure in place.

%% file: Backgroud_SIR_SEIR_Model.tex

\section{The reproduction number} \label{reproduction_number}

Infectious diseases can be analyzed with a so-called reproductive number $(R)$ that quantify the invasion or extinction of diseases in a population \cite{van2017reproduction}. 
More precisely, the $R$ is mainly used to determine the infectivity of a contagious disease.  Alternatively, we can say that $R$ represents the speed with which a disease spreads in a population. So, the primary task  is to impose policies to control the $R$ in order to control the contagious disease. This can be achieved by zonal shut down, social distancing and other factors that lead to control the outbreak of the disease in a particular geographic region. 
Fundamentally there are  two types of $R$s: (i) basic reproduction number $R_0$ and (ii) effective reproduction number $R(t)$. 
The $ R_0 $ refers to a measurement of the average number of cases that an infected person can spread over the person's infection period in a population \cite{fraser2009pandemic}. Let,  the $R_0$ of COVID-19 in a region $x$ is $y$. Then, each infection may lead to $y$ number of new secondary cases in the region $x$ i.e., each infected person may infect $y$ new individuals in the $x$ zone. 
However, $R_0$ refers to a value that can indicate that the rate of the infected population has fallen or increased or remains constant. 

Mathematically, $R_0<1$ indicates that the epidemic is in decline and it can be considered as under control. Contrarily, $R_0>1$ implies that the epidemic is on the rise and therefore cannot be considered as under control. And finally $R_0 = 1 $ demonstrates that the infection rate remains constant.

Contrarily, the $R(t)$ is used to measure the infected cases when there is a certain immunity or certain interventions are taken place. In other words, $ R (t) $ is the number of infected cases calculated  in a certain population over the period of time $ t $, taking into account that infected people are immune to infectious diseases at any given time.
Therefore, we use the actual reproduction number $ R(t)$ in order to measure the number of newly infected individuals, on average, infected by a single person at time  $t$ in a population.
In effect, $R(t)$ represents the time variant $R$ of the susceptible population where the change may be in  decline or on the rise or remains constant. These three aforementioned conditions can be expressed as  $R(t)\geq 0$.
Likewise $R_0$, $R(t)<1$, suggests the epidemic is on the decline and can be considered under control at time $t$. 
At $R(t)>1$ the epidemic is on the rise and not in  control at $t$. Finally, $R (t) = 1$ implies the infection rate remains constant.
Interestingly, $R(t)$ and $R_0$ can be related to utilizing a simple relationship as $R(t)$ can be measured with $R_0*S$ where $S$ represents the number of infected people in a particular population.
In case the immunity to the disease of a particular population is high then $S$ becomes low. Consequently, $R(t)$ becomes below 1. 
The implication is that as herd immunity is achieved, the number of new cases in the population will decrease to zero over time \cite{rodpothong2012viral}.

\begin{table*}[htb!]
\small  
  \begin{center}
    \caption{COVID-19 Dataset }
    \label{tab:dataset}
    \begin{tabular}{p{1.5cm}|c|c|c|c|c|c|c} 
    \hline
      \textbf{Country} & \textbf{\thead{Date of the \\ first case}}
       & \textbf{End date}& \textbf{\thead{Total Confirmed \\ cases}}& \textbf{\thead{Total death \\ cases}}&\textbf{\thead{Total recovered \\ cases}}&\textbf{\thead{Population}}&\textbf{\thead{Tested per \\ million people}}\\
      \hline   
      Bangladesh &2020-03-08 &2020-06-19&105535 &1388 &42945&161376708&4892
      \\
         
      India &2020-01-30&2020-06-19&395048&12948&213831&1380004385&9995\\
         
      Pakistan &2020-02-25&2020-06-19 &171666&3382&63504&220695321&6117\\
      \hline   
     
    \end{tabular}
  \end{center}
\end{table*}

In this paper, we are going to estimate $R$, i.e., $R_0$ and $R(t)$ through a variety of methods using the dataset of the selected countries such as Bangladesh, India and Pakistan. 
In the next section, we have explained the five most popular methods for estimating of $R_0$ and $ R (t) $.

\section{Epidemic forecasting models}\label{Epidemic_models}

There are numerous models that have been proposed and applied to the area of infectious COVID-19. In order to find the transmission potential in the subcontinent, we used five models, namely,  SIR model,  Exponential growth, Sequential Bayesian method, Maximum likelihood estimation, and Time-dependent estimation, briefly described in what follows.

\subsection{SIR model} \label{sirm}
The SIR model is a basic mathematical model for describing the dynamics of infectious diseases. It is also called the \emph{compartmental model} because the model divides the population into different compartments. More particularly, the population of size $N$ is divided into three compartments \cite{calafiore2020modified}: \emph{{\bf S}usceptible}, \emph{{\bf I}nfectious}, and \emph{{\bf R}ecovered}, which will be detailed in what follows.

{\bf Susceptible:} is the number of people who are vulnerable to exposure with infectious people around at time $t$, denoted by $S(t)$.

{\bf Infectious:} is a group of people who are infected with the disease. Moreover, they can spread the disease to susceptible people and can be recovered from it, in a specific time $t$, denoted by $I(t)$. 

{\bf Recovered:} is a number of people who get immunity in a time $t$, denoted by $R'(t)$. Therefore, they are not susceptible to the same disease anymore. 

However, we can write the SIR model as a differential equation of each compartment \cite{calafiore2020modified}: 

\begin{equation}
\frac{\partial s}{\partial t} = - \beta *s(t)*i(t)
\end{equation}
\begin{equation}
\frac{\partial i}{\partial t} = \beta *s(t)*i(t) - \gamma *i(t)
\end{equation}
\begin{equation}
\frac{\partial r}{\partial t} = \gamma *i(t)
\end{equation}

where $t$ defines the time, $s(t)=\frac{S(t)}{N}$, $i(t)=\frac{I(t)}{N}$, $r(t)=\frac{R'(t)}{N}$, and $N=(S(t)+I(t)+R'(t))$. Likewise, $\beta$ is a controlling parameter that defines the number of people infected by exposure in a specific time $t$, and $\gamma$  defines the ratio of the infected individuals who can recover in a time $t$. Using these two parameters (i.e., $\beta$, $\gamma$) we can estimate the $(R_0)$, mathematically, $R_0=\frac{\beta}{\gamma}$, which defines the average number of people infected from single disease exposure. Hence, if the $R_0$ value is higher, the probability of the pandemic is also higher. 


\subsection{Exponential growth (EG)}
The exponential growth (EG) rate is an important measure to see the speed of the spread of an infectious disease. As introduced in section \ref{sirm}, the exponential growth rate can be written as $ r = \beta - \gamma$, where, $\beta$ defines the number of people infected by exposure in a specific time $t$, and $\gamma$  defines the ratio of the infected individuals who may recover in time $t$. 
However, $r$ is a disease threshold value when $r$ is at zero (i.e., $r=0$). If $r$ is positive (i.e., $r > 0$), the disease can invade a population, whereas it cannot invade a population, if $r$ is negative (i.e., $r < 0$) \cite{ma2020estimating}.
The relationship between $R_0 $ and growth rate $r $ is not simple. For a specific distribution over generation time (e.g. gamma distribution), it can sometimes be simplified.
Assuming that the generation time is completely constant, such as $T$, the $R$ can be written as \cite{roberts2007model}, $R_0 = e ^ {(r * T)} $.

\subsection{Sequential bayesian method (SB)} The sequential bayesian approach can be used to estimate the initial reproduction number ($R_0$). Technically, the Bayesian method works in the context of probabilistic modeling. Therefore, the probability model of $R_0$ can be written as follows \cite{bettencourt2008real}. 
\begin{align*}
P[R_0]=\frac{P[R_0|\nabla T(t+\tau)\leftarrow \nabla T(t)]*P[\nabla T(t+\tau)\leftarrow \nabla T(t)]}{P[\nabla T(t+\tau)\leftarrow \nabla T(t)|R_0]}
\end{align*}
where, $T(t)$ is the total number of cases up to time $t$, and the occurrence of new infected cases over the period $\tau$, and $\nabla T(t+\tau) = T(t+\tau)-T(t)$. The probability distribution $P[\nabla T(t+\tau)\leftarrow \nabla T(t)]$ and $P[\nabla T(t+\tau)\leftarrow \nabla T(t)|R_0]$ are independent and dependent on $R_0$, respectively. Hence, the sequential bayesian estimation of $R_0$ can be made using the posterior distribution for $R_0$, at time $t$ as the prior in the next estimation step at time $t+\tau$. 

\subsection{Maximum likelihood estimation (ML)} The maximum likelihood-based estimation relies on two considerations : (i) the number of secondary cases produced by an infected individual follows a Poisson distribution, and (ii) the expected value $R_0$. Suppose the data is a periodic incidence e.g., \{$N_0$, $N_1$,... $N_t$\}, where, $t$ is a time unit, and $N_t$ defines the number of new cases at time $t$. Therefore, the maximum likelihood can be estimated as follows \cite{forsberg2008likelihood}: $ML(R_0)=\displaystyle \prod_{t=1}^{T} \frac{\exp^{-\mu_t}\mu_t^{N_t}}{N_t!}$, where, $\mu_t=R_0 \displaystyle \sum^{min(k,t)}_{j=1}N_{t-j}*w_j$, where $k$ is the constraint and $w_j$ is the time distribution. Here, $k<T$. Note that the maximum likelihood-based method is used for the estimation of the basic reproductive number ($R_0$). 

\subsection{Time dependent estimation (TD)} Typically, the time-dependent $R$ estimation is not straightforward, because we can only see the epidemic curve. There is no information about who infected whom. However, using likelihood-based estimates of $R$,  the time dependant $R$ can be written as \cite{obadia2012r0}: $R_t=\frac{1}{N_t} \displaystyle \sum_{t_j=t}R_{j}$, where $N_t$ specifies the number of new cases at time $t$, and $R_{j}$ is the $R(t)$ for case $j$, that is the sum over all cases $i$,  mathematically, $R_{j}=\displaystyle \sum_{i}p_{ij}$, where $p_{ij}$ defines the relative likelihood. More precisely, the relative likelihood that case $i$ has been infected by case $j$, normalized by the likelihood that case $i$ has been infected by any other case $k$, which can be written as follows \cite{wallinga2004different}: $p_{ij}=\frac{w(t_i-t_j)}{\sum_{i \neq k}w(t_i-t_k)}$, where, $w(t_i-t_j)$ defines the time interval of the infectious diseases.

%% file: Dataset.tex
\section{Data source}\label{data}

We use a publicly available COVID-19 dataset\footnote{https://github.com/RamiKrispin/coronavirus} extracted from the coronavirus repository of the Johns Hopkins University Center for Systems Science and Engineering (JHU CCSE). 
This dataset contains daily observations on COVID-19 confirmed, recovered and death cases for most countries over the world.
More precisely, the incidence data are provided on a daily basis.
For this analysis, we consider data from South Asian countries such as Bangladesh, India, Pakistan, Nepal, Bhutan, Maldives and Sri Lanka. Nonetheless, the top three populated countries (Bangladesh, India, and Pakistan) finally have been selected for this particular article. Data dated upto June 19, 2020 is utilized. 
The entire dataset of COVID-19 cases of these selected countries is observed for each day, as shown in Figure \ref {fig_con_dea_rec}. In addition, the Table \ref{tab:dataset} sequentially displays the date of the first confirmed case, the total number of confirmed incidences, the total number of deaths, the total number of recovery cases, population and COVID-19 tested per million people in each country.

The surprising fact is that these countries have conducted very few COVID-19 tests in susceptible populations. Therefore, it is very likely that this data set may not cover the real scenario of the COVID-19 situation. However, we can use this data set to mainly calculate the reproduction numbers, so as to observe the spread of COVID-19.

%% file: Experiments.tex
\section{Experimental results}\label{experimental_results}
We conduct several experiments to analyze the trend of COVID-19 in the concerned countries. 
The first experiment focuses on COVID-19 cases as confirmed, deceased and recovered to compare the COVID-19 situation in these countries. The motivation is to understand the deteriorated conditions for the COVID-19 pandemic in this region.
Next, we conduct experiments through SIR, EG, SB, ML and TD models to estimate $R$s for these countries and predict the COVID-19 pandemic. The result is analyzed to observe the model performances on the COVID-19 dataset. 
For this, we use the R0 package provided by R programming language to impletement the code for the above methods \cite{obadia2012r0}. 
More particularly, we employ \textit{estimate.R} function of R0 package to apply the above methods to a given epidemic curve.
The result is analyzed to observe the model performances on the COVID-19 dataset. 

\begin{figure}[htb!]
\centering
\captionsetup{justification=centering}
\includegraphics[scale=0.30]{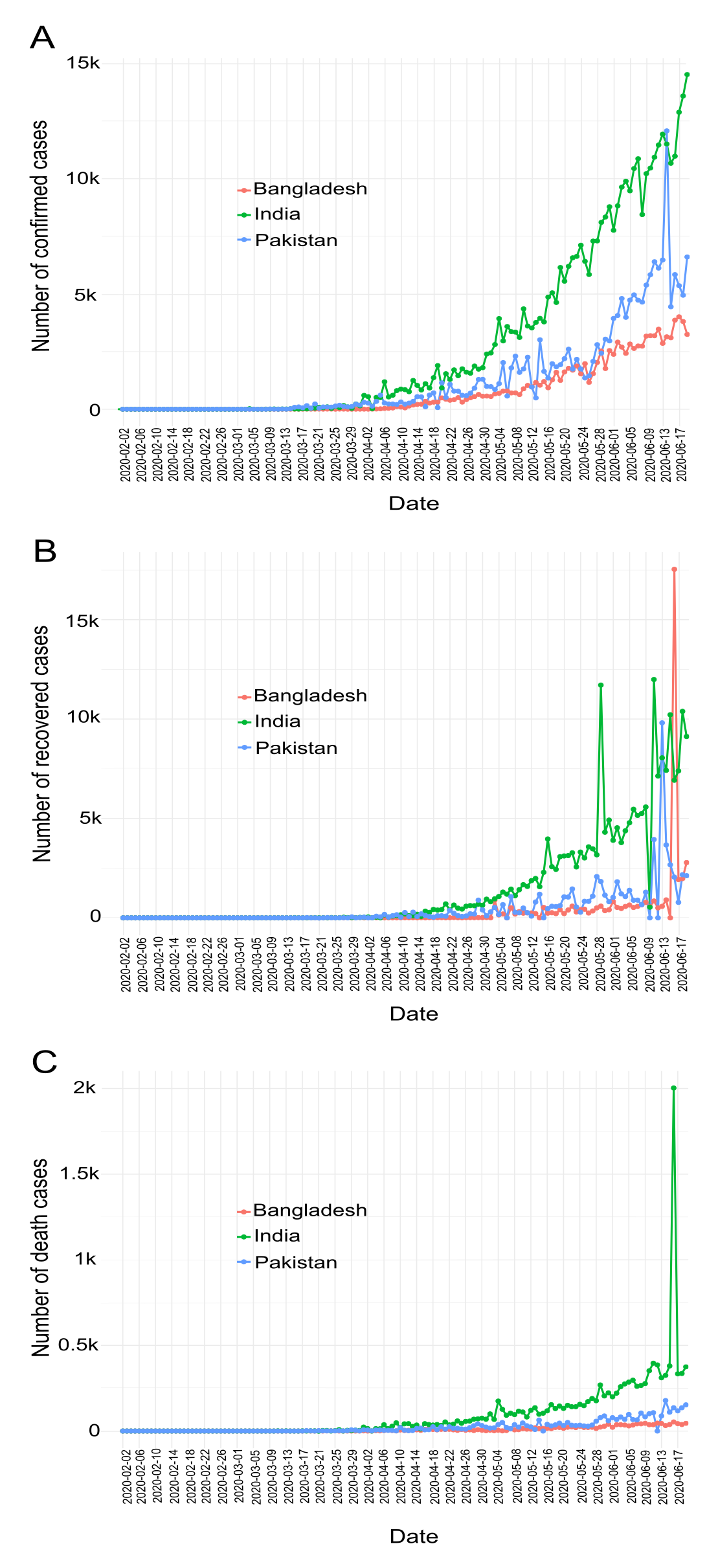}
\caption{Daily COVID-19 cases in Bangladesh, India and Pakistan: A) Confirmed cases B) Recovered cases C) Death cases }
   \label{fig_con_dea_rec}
\end{figure}

\begin{figure}[htb!]
\centering
\captionsetup{justification=centering}
\includegraphics[scale=0.30]{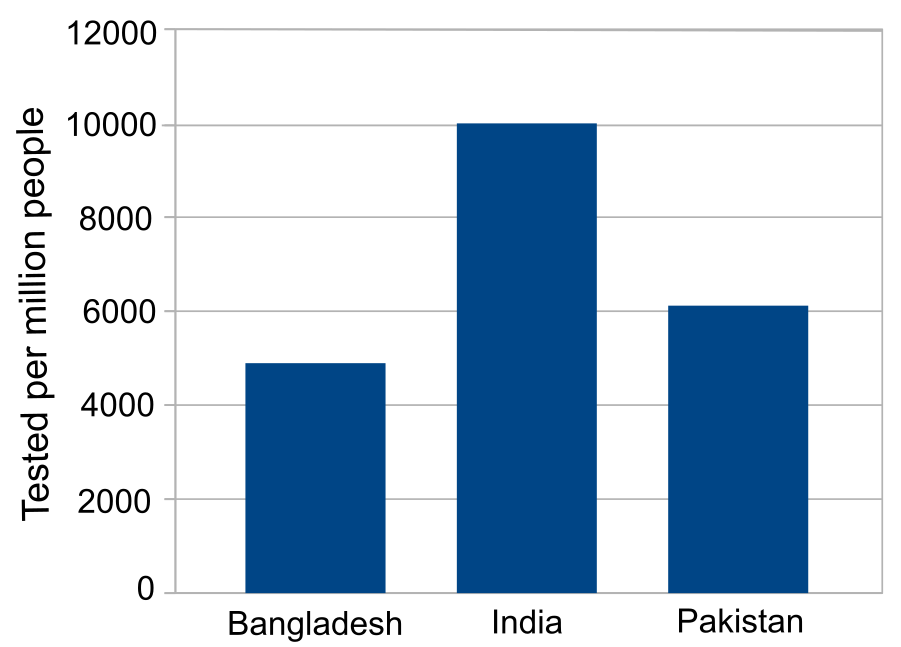}
\caption{COVID-19 cases tested per million people in Bangladesh, India and Pakistan}
\label{fig:tested_per_million}
\end{figure}

\begin{figure}[htb!]
\centering
\captionsetup{justification=justified}
\includegraphics[scale=0.30]{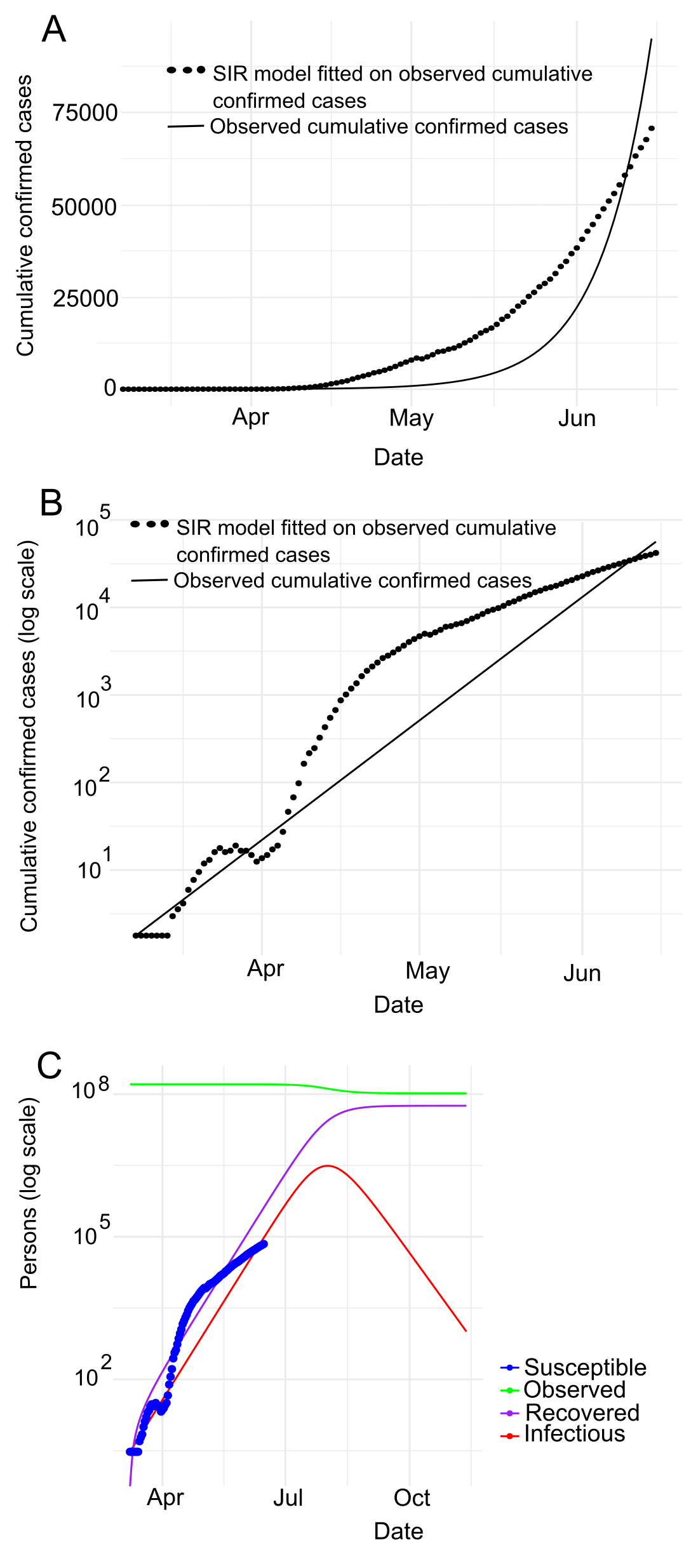}
\caption{Prediction with SIR model for Bangladesh: A) SIR model fitted on observed cumulative infected cases B) SIR model fitted on observed cumulative infected cases with semi log scale C) SIR model prediction with no human interaction}
   \label{BD_fig:1-2-3}
\end{figure}

\begin{figure}[htb!]
\centering
\captionsetup{justification=justified}
\includegraphics[scale=0.30]{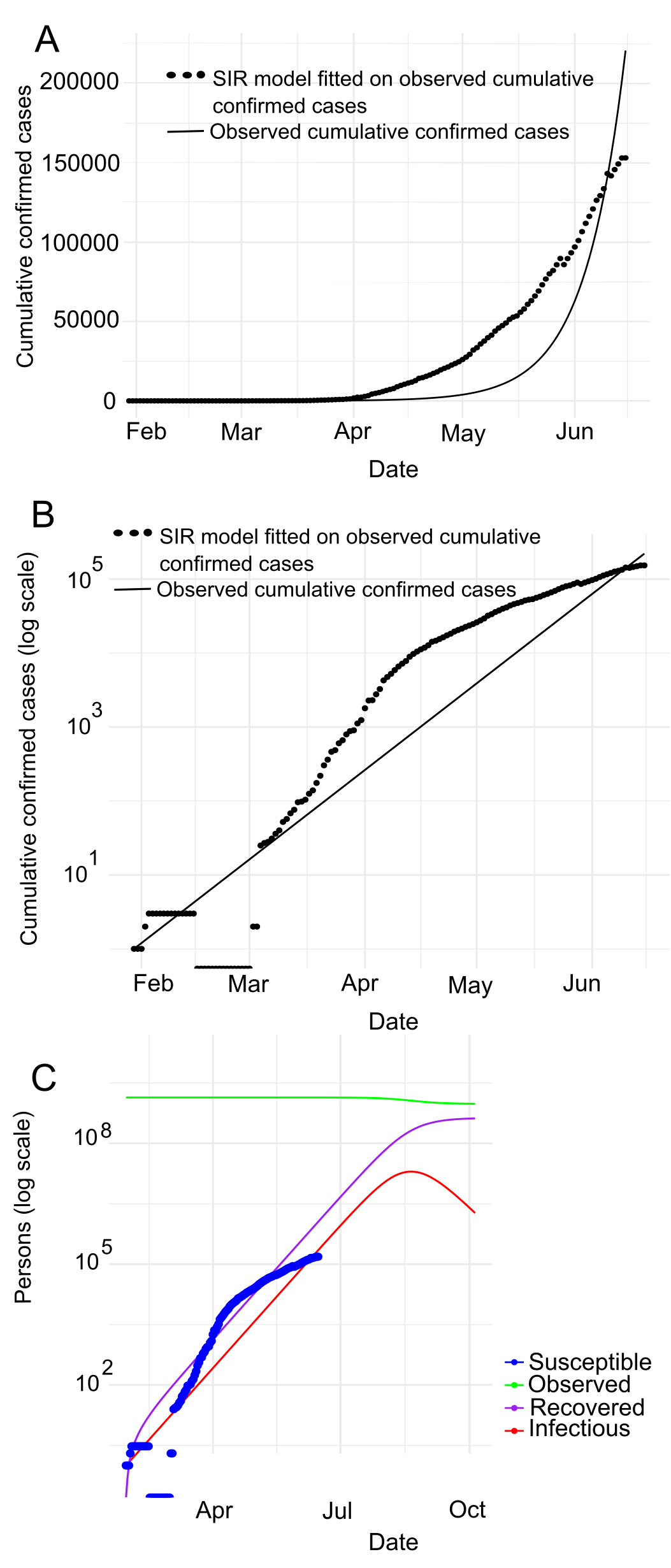}
\caption{Prediction with SIR model for India: A) SIR model fitted on observed cumulative infected cases B) SIR model fitted on observed cumulative infected cases with semi log scale C) SIR model prediction with no human interaction}
   \label{Ind_fig:1-2-3}
\end{figure}

\begin{figure}[htb!]
\centering
\captionsetup{justification=justified}
\includegraphics[scale=0.30]{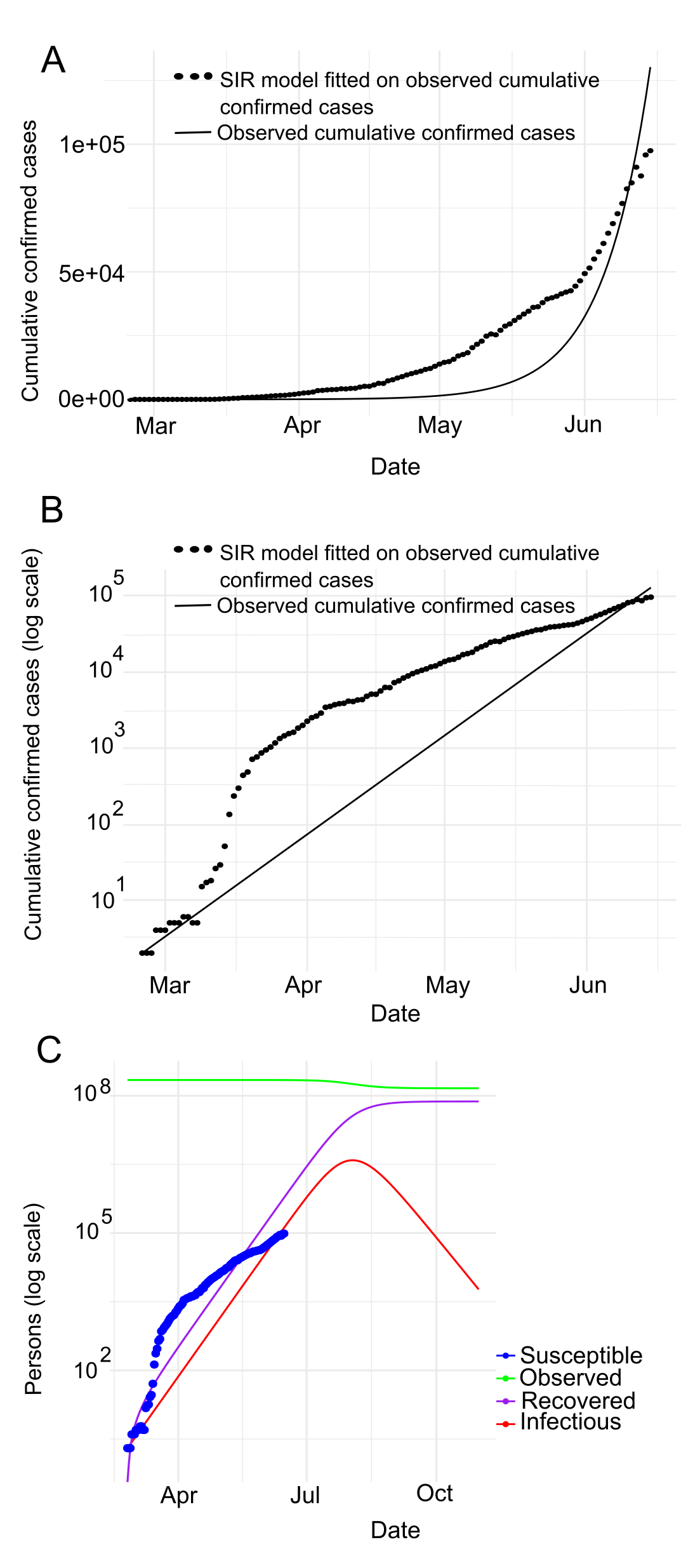}
\caption{Prediction with SIR model for Pakistan: A) SIR model fitted on observed cumulative infected cases B) SIR model fitted on observed cumulative infected cases with semi log scale C) SIR model prediction with no human interaction}
   \label{Pak_fig:1-2-3}
\end{figure}

\subsection{COVID-19 cases}

This subsection analyzes the COVID-19 data in order to observe confirmed, death and recovered cases in Bangladesh, India and Pakistan. Figure \ref{fig_con_dea_rec} shows the number of confirmed (Figure \ref{fig_con_dea_rec}A), recovered (Figure \ref{fig_con_dea_rec}B), and death (Figure \ref{fig_con_dea_rec}C) cases in Bangladesh, India and Pakistan respectively. The Figure depicts COVID-19 cases in all these countries are increasing every day. In addition Figure \ref{fig_con_dea_rec}, presents that the spread of COVID-19 in India exceeds that of Bangladesh and Pakistan. The Figure also illustrates that  Bangladesh has less confirmed cases and deaths apparently more control over the spread of COVID-19 than India and Pakistan as it. Intuitively, the citizens of Bangladesh comparatively better maintained medical care, lock-down and social distances, as a result COVID-19 spread slower from person to person compared to the other two countries and vice versa. 

However the aforementioned argument is inconclusive and rather unlikely as Bangladesh conducted the least amount of tests per person compared to the other two. 
Figure \ref{fig:tested_per_million} shows that SARS-CoV-2 tests per million people performed in Bangladesh, Pakistan and India are 4892, 6117 and 9995 respectively. 
Clearly, more tests reveal more infected people. With unexpectedly poor and nonuniform samples we rather not conclude as above and investigate further and focus on how COVID-19 is spreading in the subcontinent. Intuitively, the subcontinent case may reflect COVID-19 spreading in the developing countries, especially spreading in South Asia.

Therefore, the further analysis focuses on the transmission speed of COVID-19 using different methods.

\begin{table*}[h]
  \begin{center}
  \small 
    \caption{Prediction with SIR model }
    \label{tab:table1}
    \begin{tabular}{l|c|c|c} 
    \hline
      \textbf{The predicted values for the following parameters} & \textbf{Bangladesh} & \textbf{India}& \textbf{Pakistan}\\
      \hline
      Infection Rate, $\beta$ & 0.5524  &0.5449    & 0.550  \\
      Recovery rate, $\gamma$ & 0.4475  &0.4550 & 0.449 \\
      $R_{0}=\dfrac{\beta}{\gamma}$ & 1.234 &1.197  &1.22\\
      
     Herd immunity threshold ($1- \dfrac{1}{R_0})\times 100\%$  &18.97 \% of population & 16.49\% of population&18.18\% of population\\ 
      Peak of Pandemic  &2020-08-01  &2020-08-20 & 2020-08-03\\
      Maximum Infected &3109321 &19884176&3891427 \\
     Severe cases (assume 20\% of Infected cases)    &621864  &3976835 &  778285\\
     Patients need intensive care (assume 6\% of Infected cases) &186560&1193051&233485 \\
     Deaths assumed for 3.5\% fatality rate &108826& 695946&136200\\
     \hline
    \end{tabular}
  \end{center}
\end{table*}

\subsection{Prediction with SIR model}

In this experiment, we use the SIR model to predict COVID-19 cases in Bangladesh, India and Pakistan.

{\bf{Bangladesh.}}
Figure \ref{BD_fig:1-2-3}A depicts the SIR model fitting to the number of observed confirmed cases where Figure \ref{BD_fig:1-2-3}B presents the same observation in the logarithmic scale in Bangladesh.
We observe that the number of cases (black dotted line as shown in Figure \ref{BD_fig:1-2-3}A and in Figure \ref{BD_fig:1-2-3}B) follows the number of confirmed cases expected (black line as shown in Figure \ref{BD_fig:1-2-3}A and Figure \ref{BD_fig:1-2-3}B) by SIR model. 
Note that, the observed data and predicted values overlapping with each other indicates COVID-19 clearly is in an exponential phase in Bangladesh. 
Figure \ref{BD_fig:1-2-3}B shows that the curve is flattening in between mid  March (around) to 1st April. This impliedly indicates that the spreading of COVID-19 is comparatively in control in Bangladesh. 
Furthermore, the slope of the curve dropping down in between mid-March and April 1st conforms to the aforesaid observation (spreading of COVID-19 is in control in Bangladesh during this period).
But then, the slope is going up at the steepest of all times until the end of May, i.e. in this time COVID-19 is spreading rapidly. And then, the slope remains constant with a comparatively lower value.

Furthermore, we carry out the experiment to adapt the SIR model to confirmed, death and recovered cases, as shown in Figure \ref{BD_fig:1-2-3}C. With this experiment, we derive the estimated values of various parameters as shown in Table \ref{tab:table1}. According to Table \ref{tab:table1}, we can see that the $R_0$ is around 1.23 indicating that COVID-19 is spreading in Bangladesh day by day. Based on the $R$, we calculate the herd immunity threshold using the equation $ 1- \dfrac {1} {R_0} $~\cite {fine2011herd}, i.e., $18.97 \%$. With this trend, the epidemic will be at its peak in 2020-08-01. The SIR model estimates that the maximum infected population in Bangladesh will be $3109,321$. Of these, the serious cases will be $621,864$ (assuming $20\%$ of the infected population). It also shows that around $186,560$ (assuming $6\%$ of the infected population) people need intensive care and up to $108,826$ deaths (assuming $3.5\%$ mortality rate). 


\begin{figure*}[htb!]
	\centering 
	\begin{subfigure}[b]{1\columnwidth}
		\includegraphics[width=0.950\linewidth]{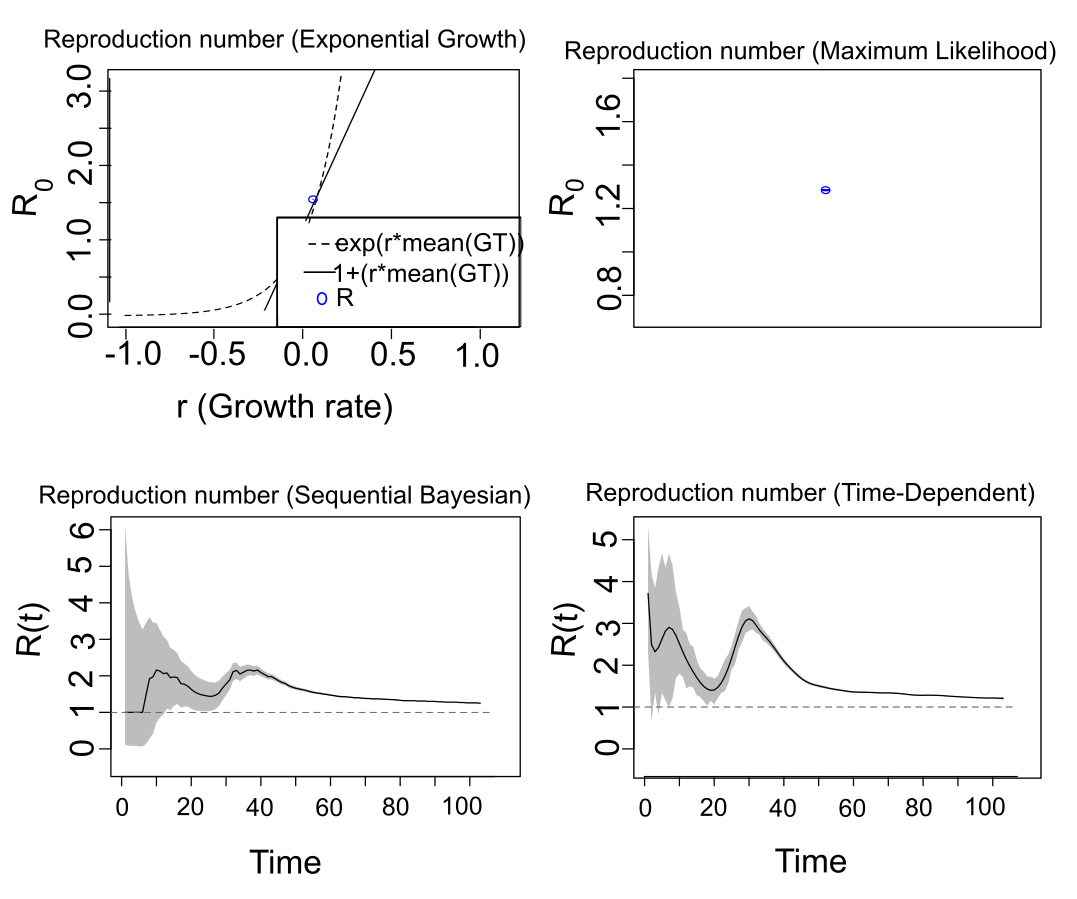}
		\caption{Bangladesh}
		\label{Ban_R0/R(t)}
	\end{subfigure}
	\par
	
	\begin{subfigure}[b]{1\columnwidth}
		\includegraphics[width=.950\linewidth]{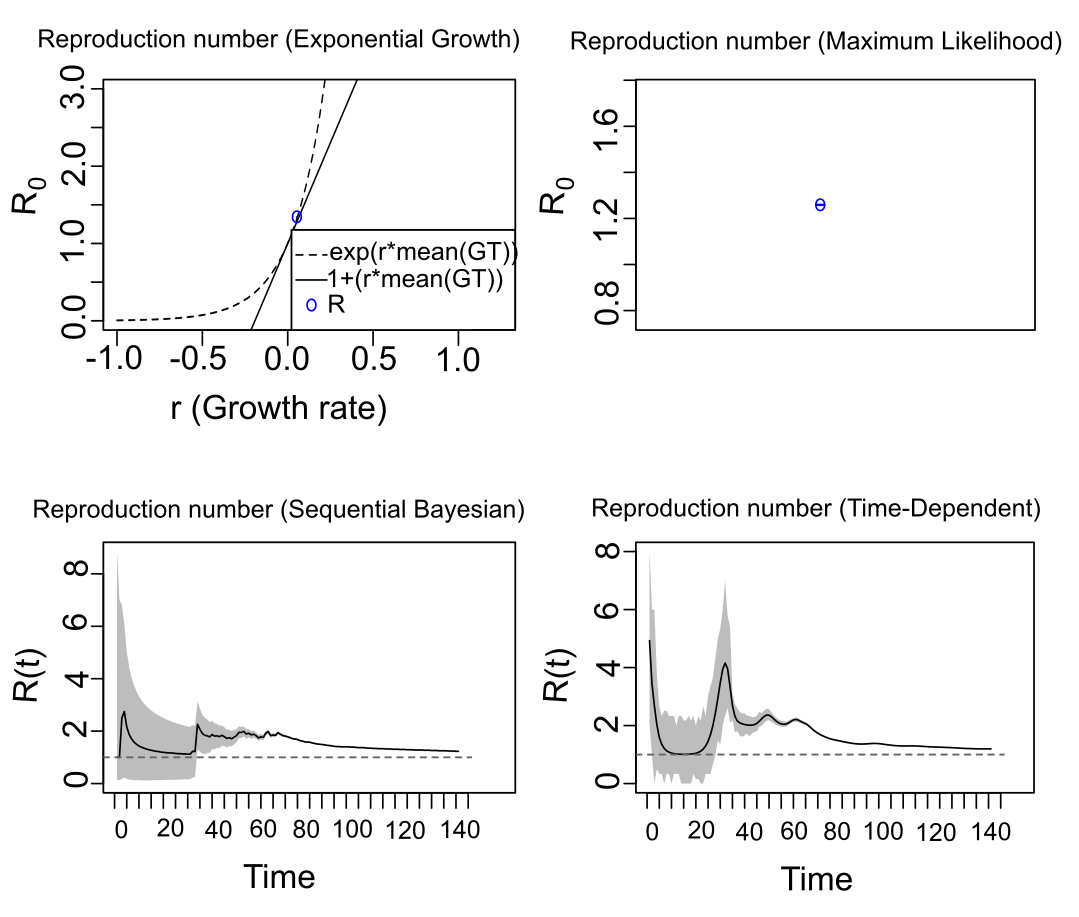}
		\caption{India}
		\label{Ind_R0/R(t)}
	\end{subfigure}
	\hfill
	\begin{subfigure}[b]{1\columnwidth}
		\includegraphics[width=.950\linewidth]{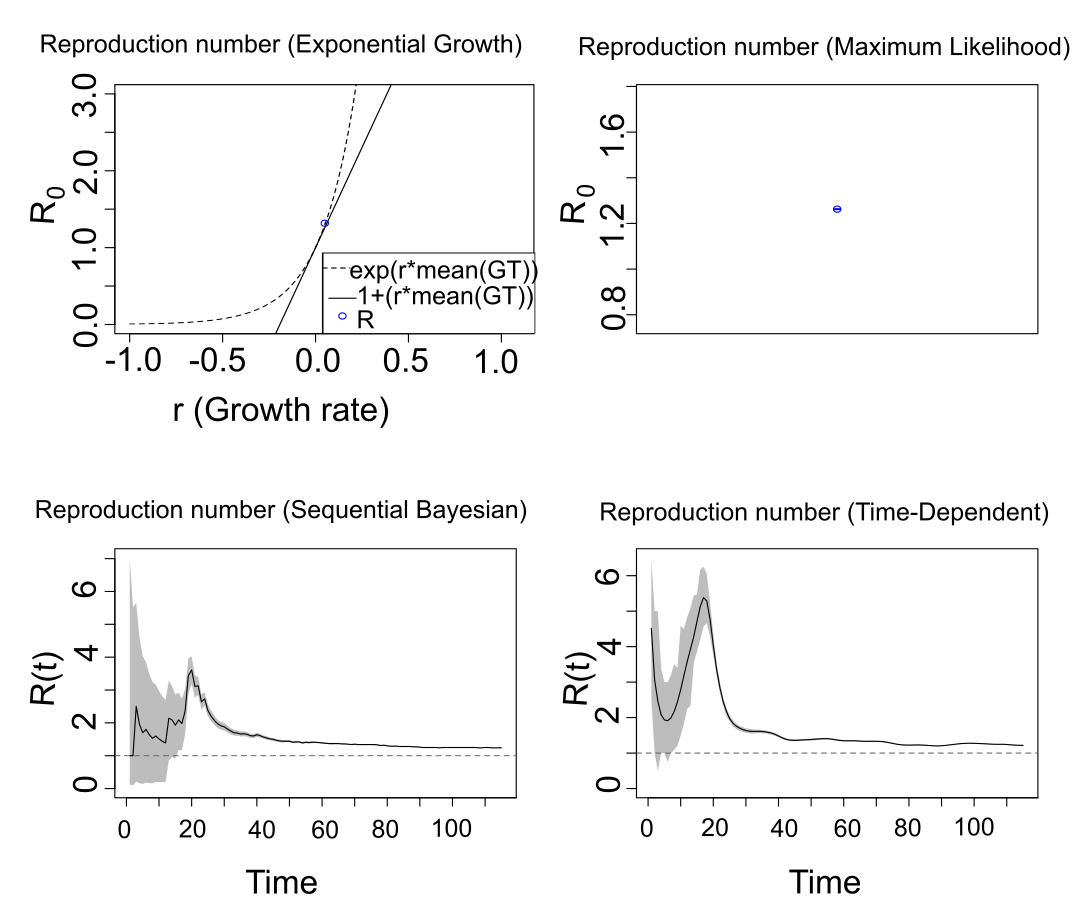}
		\caption{Pakistan}
		\label{Pak_R0/R(t)}
	\end{subfigure}
	\caption{Estimate of the reproduction number $R_0/R(t)$ in Bangladesh, India and Pakistan}
	\label{fig:R0(Rt)}
\end{figure*}


\begin{table*}[htb!]
\centering
\caption{ $R_0/R(t)$ estimation by different methods} 
\label{tab:R0/R(t)} 
\resizebox{\linewidth}{!}{\begin{tabular}{|p{2.9cm}|c|p{3.4cm}|p{3.4cm}|c|c}\hline
\multicolumn{1}{p{2cm}|}{\multirow{1}{*}{Methods}} &\multicolumn{1}{c|}{SIR}&\multicolumn{1}{c|}{EG}&\multicolumn{1}{c|}{ML} &\multicolumn{1}{c|}{TD} &\multicolumn{1}{c}{SB}\\\cline{1-6}
\multicolumn{1}{p{3cm}|}{\multirow{1}{*}{Reproduction number}}&$R_0$&$R_0$ [CI.lower, CI.upper]&$R_0$ [CI.lower, CI.upper]&R\textsubscript{mean}(t)  [R\textsubscript{low}(t), R\textsubscript{high}(t)]&R\textsubscript{mean}(t)[R\textsubscript{low}(t), R\textsubscript{high}(t)]\\

\multicolumn{1}{p{2cm}|}{\multirow{1}{*}{Bangladesh}} &1.234 & 1.380 [1.380, 1.381] & 1.288 [1.286, 1.290]  & 1.746 [1.209, 3.715]& 1.555 [1.000, 2.16]\\

\multicolumn{1}{p{2cm}|}{\multirow{1}{*}{India}} &1.197& 1.344 [1.344, 1.344] &1.259 [1.257, 1.260] & 1.668 [1.007, 4.928] & 1.507 [1.00, 2.75] \\

\multicolumn{1}{p{2cm}|}{\multirow{1}{*}{Pakistan}} & 1.220 & 1.319 [1.318, 1.319] & 1.264 [1.262, 1.266] & 1.774 [1.202, 5.381] & 1.560 [1.00, 3.61] \\
\hline

\end{tabular}}
\end{table*}

{\bf{India.}}
Figure \ref{Ind_fig:1-2-3}A and Figure \ref{Ind_fig:1-2-3}B represents the cumulative COVID-19 infected cases in number and in logarithmic forms respectively in India. It clearly depicts that the spread of COVID-19 is in an exponential phase in this particular period.

In addition, Figure \ref{Ind_fig:1-2-3}B shows that the slope of the curve (black dotted line) was not steep in between February 1 and mid-March, indicating that the spread of COVID-19 in India was under control during this period. But then, it rose almost exponentially, which means that this time COVID-19 is spreading rapidly.
Furthermore, Figure \ref{Ind_fig:1-2-3}C depicts the experimental results of the SIR model for confirmed, death and recovered cases. We put all the estimated values of all the different parameters calculated using the SIR model in the Table \ref{tab:table1}.
The $R_0$ (derived from the  SIR model is about $1.197$) indicates that COVID-19 accelerates over time, where the computed herd immunity is $16.97\%$.
With this $R_0$ the epidemic will peak in 2020-08-20.
Moreover, according to the SIR model, the maximum infected population is $19,884,176$, among them, the serious cases will be $3,976,835$ (assuming $20\%$ of the infected population). Furthermore, around $1,193,051$ (assuming $6\%$ of the infected population) people need intensive care and up to $695,946$ deaths (assuming a mortality rate of $3.5\%$) (see Table \ref{tab:table1}).

{\bf{Pakistan.}}
Like Bangladesh and India, we have conducted experiments on Pakistan's COVID-19 data through the SIR model.
Figure \ref{Pak_fig:1-2-3}A and Figure \ref{Pak_fig:1-2-3}B depicts the observed cumulative infected incidence in cases and in cases in logarithmic scale respectively. The figures indicate that the COVID-19 is spreading exponentially.
In addition, Figure \ref{Pak_fig:1-2-3}B shows that the slope of the curve (black dotted line) was not accentuated till mid-March, indicating that the spread of COVID-19 in Pakistan. But then it almost shot up, i.e., the COVID-19 is spreading rapidly during this time period.
In order to observe deeper insights, we have conducted SIR model experiments on confirmed, death, and recovered cases, as shown in \ref {Pak_fig:1-2-3}C.
The estimated values of all the various parameters of the SIR model are shown in Table \ref{tab:table1}. It is worth noting that the estimated $R_0$ of the SIR model is about $1.22$, indicating that COVID-19 accelerates over time. 
We find Pakistan's herd immunity as $18.18\%$. The epidemic will be at its peak in 2020-08-03 with $ R_0 $ remaining at this particular rate. 
According to the SIR model the maximum number of infection cases is $3,891,427$. Among them, the serious cases are $778,285$ (assuming $20\%$ of the infected population), the intensive care requirement is $233,485$ approximately (assuming $6\%$ of the infected population) and estimated maximum deaths is $136,200$ (assuming a mortality rate of $3.5\%$).

{\bf{Comparison.}}
Table \ref{tab:table1} shows all the predicted values of various parameters of Bangladesh, India and Pakistan estimated by the SIR model. It is worthy to mention that the $R_0$ of Bangladesh and Pakistan are very close, i.e., $1.22$ and $1.23$, respectively, while $R_0$ of India is about $1.19$. 
It concludes that Bangladesh and Pakistan are experiencing an increasing number of infected people compared to India. In other words, India has so far somewhat stronger control over the spread of COVID-19 compared to Bangladesh and Pakistan.

Next, we will use the rest of the methods (i.e., EG, SB, ML and TD) to estimate the $R$ in order to verify what the SIR model foresees is consistent with other models.


\begin{figure*}[htb!]
	\centering 
	\begin{subfigure}[b]{1\columnwidth}
		\includegraphics[width=0.95\linewidth]{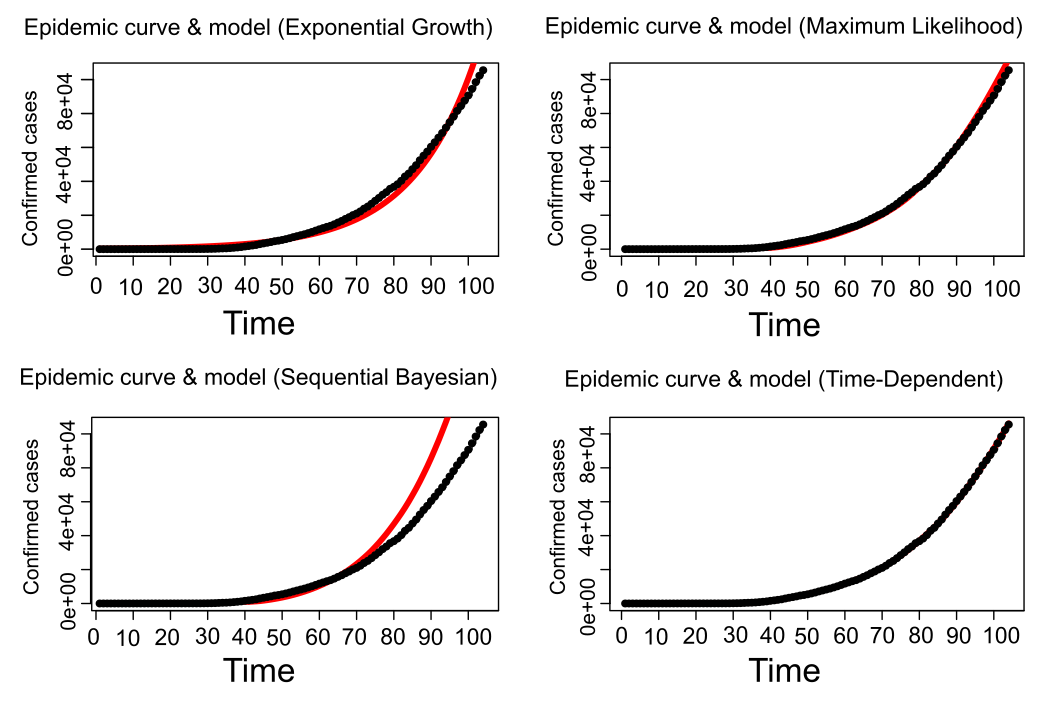}
		\caption{Bangladesh}
		\label{Ban_pre}
	\end{subfigure}
	\par
	
	\begin{subfigure}[b]{1\columnwidth}
		\includegraphics[width=0.95\linewidth]{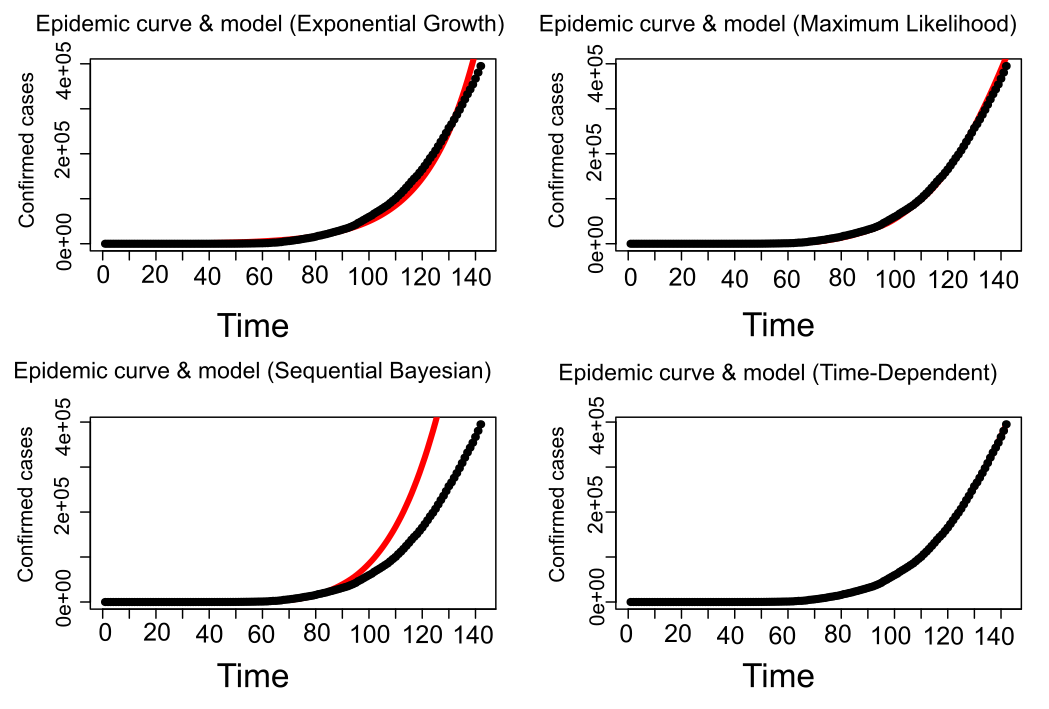}
		\caption{India}
		\label{Ind_pre}
	\end{subfigure}
	\hfill
	\begin{subfigure}[b]{1\columnwidth}
		\includegraphics[width=0.95\linewidth]{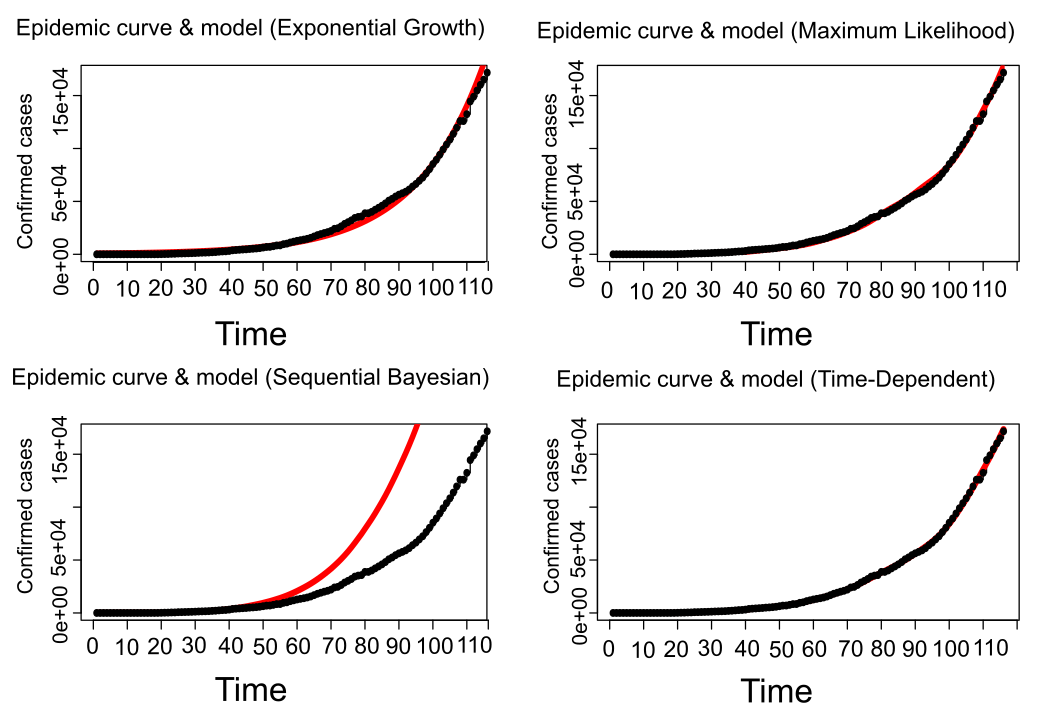}
		\caption{Pakistan}
		\label{Pak_pre}
	\end{subfigure}
	\caption{Models (EG, ML, SB, TD) fitting on COVID-19 confirmed cases}
	\label{fig:prediction}
\end{figure*}

\subsection{COVID-19 reproduction number ($ R_0 $ / $R(t)$) estimation }

In this experiment, we consider various methods to compute the value of the $R_0$ and $R(t)$. This experiment investigates how the $R$ has changed with the various methods of the estimation process, the role of excessive dispersion in the distribution of secondary cases and the aggregation of the epidemic curve at ever larger time intervals. In this experiment, we assume that the sequence interval of COVID-19 for Bangladesh, India and Pakistan is equal to the sequence interval of COVID-19 in Wuhan, China with a mean of 5.2 days and a standard deviation of 2.8 days \cite{ganyani2020estimating}.

Figure \ref{fig:R0(Rt)} depicts an estimation of $R_0$ and $R(t)$ for Bangladesh (see Figure \ref{Ban_R0/R(t)}) in the period from March 08 to June 19 (for 104 days), for India (see Figure \ref{Ind_R0/R(t)}) in the period from January 30 to June 19 (for 142 days), and for Pakistan (see Figure \ref{Pak_R0/R(t)}) in the period from February 25 to June 19 (for 116 days), respectively. Moreover, Table \ref{tab:R0/R(t)} reports the computed $ R_0 $ and $R(t)$ for Bangladesh, India and Pakistan which we have estimated with various methods. 

According to the Table \ref{tab:R0/R(t)}, the $R_0$ for Bangladesh generated with the methods of exponential growth (EG), maximum likelihood (ML) are $1.3809$ [$1.3803$, $1.3815$] and $1.2884$ [$1.2860$ $1.2908$] respectively which is higher than that of India i.e., $1.3446$ [$1.3443$, $1.34448$] with EG and $1.2591$ [$1.2579$, $1.2602$] with ML and Pakistan $1.3192$ [$1.3189$, $1.3196$] with EG and $1.2645$ [$1.2626$, $1.2662$] with ML. Moreover, Table \ref{tab:R0/R(t)} also shows the estimated values of $R(t)$ for Bangladesh, India and Pakistan using the methods of sequential Bayesian and time-dependent. The results show that $R(t)$ value for Pakistan is higher than India and Bangladesh.
According to the obtained results depicted in Table \ref{tab:R0/R(t)}, we have observed that for different methods the $R$ does not follow the same trend of values. Therefore, we proceed to observe the curve fitting using these $R$s on the COVID-19 data.
Figure \ref{fig:prediction} shows that the $R$ generated by maximum likelihood (ML) and time-dependent methods fit well to the data compared to the other methods for all countries. 
Moreover, Figure \ref{fig:prediction} shows that SB method fits very poorly to the data for Bangladesh, India and Pakistan.
The results show that Bangladesh and Pakistan have comparatively worse control over the spread of COVID-19, meaning $ R (t) $ is getting higher than that of India.
However, it is worth noting that the $R_0$ and $R(t)$ in each country are greater than 1, \textit{i.e.}, COVID-19 is still spreading in all of these countries.

%% file: Conclusion.tex
\section{Conclusion}\label{conclustion}

In this article, we provide an assessment of COVID-19
outbreak and measure the spread rate among Bangladesh, India and Pakistan. We utilize the SIR model to predict important parameters such as infection rate, recovery rate, herd immunity threshold, peak of the pandemic, maximum infected, severe cases, patients needing intensive care, deaths of COVID-19 pandemics. We utilize EG, SB, ML and TD models to validate the effectiveness of the estimated parameters of the SIR model. 
For doing so, we find the basic reproduction number $R_0$ and effective reproduction number $R(t)$. 

Experiments show that in all the considering countries, the estimated effective reproduction number $R(t)$ is much larger than the basic reproduction number $R_0$, which means that the containment measures implemented by Bangladesh, India and Pakistan are ineffective and inefficient. Besides that, according to different models the reproduction numbers of Bangladesh, India and Pakistan are all higher than about $1.2$, indicating that the outbreak of COVID-19 is spreading rapidly.

Result directed recommendations are (i) to adopt stricter prevention and control measures, (ii) to improve the country's quarantine measures, define outcomes slowing down the spread of COVID-19. Failure in doing so the pandemic situation of the region may decline rapidly.

%% file: elsarticle-template-harv.bbl
\begin{thebibliography}{10}
\expandafter\ifx\csname url\endcsname\relax
  \def\url#1{\texttt{#1}}\fi
\expandafter\ifx\csname urlprefix\endcsname\relax\def\urlprefix{URL }\fi
\expandafter\ifx\csname href\endcsname\relax
  \def\href#1#2{#2} \def\path#1{#1}\fi

\bibitem{who2020pneumonia}
WHO, Pneumonia of unknown cause - china,
  \url{https://www.who.int/csr/don/05-january-2020-pneumonia-of-unkown-cause-china/en},
  [Online; accessed 30-June-2020] (2020).

\bibitem{wu2020new}
F.~Wu, S.~Zhao, B.~Yu, Y.-M. Chen, W.~Wang, Z.-G. Song, Y.~Hu, Z.-W. Tao, J.-H.
  Tian, Y.-Y. Pei, et~al., A new coronavirus associated with human respiratory
  disease in china, Nature 579~(7798) (2020) 265--269.

\bibitem{WHO2020nCoV}
WHO, Novel coronavirus - china,
  \url{https://www.who.int/csr/don/12-january-2020-novel-coronavirus-china/en},
  [Online; accessed 30-June-2020] (2020).

\bibitem{ICTV2020species}
C.~S.~G. of~the International Committee on Taxonomy~of Viruses, The species
  severe acute respiratory syndrome-related coronavirus: classifying 2019-ncov
  and naming it sars-cov-2, Nature Microbiology 5~(4) (2020) 536--544.

\bibitem{WHO2020naming}
WHO, Naming the coronavirus disease (covid-19) and the virus that causes it,
  \url{https://www.who.int/emergencies/diseases/novel-coronavirus-2019/technical-guidance/naming-the-coronavirus-disease-(covid-2019)-and-the-virus-that-causes-it},
  [Online; accessed 30-June-2020] (2020).

\bibitem{WHO2020emergency}
WHO, Statement on the second meeting of the international health regulations
  (2005) emergency committee regarding the outbreak of novel coronavirus
  (2019-ncov),
  \url{https://www.who.int/news-room/detail/30-01-2020-statement-on-the-second-meeting-of-the-international-health-regulations-(2005)-emergency-committee-regarding-the-outbreak-of-novel-coronavirus-(2019-ncov)},
  [Online; accessed 30-June-2020] (2020).

\bibitem{WHO2020pandemic}
WHO, Who director-general's opening remarks at the media briefing on covid-19 -
  11 march 2020,
  \url{https://www.who.int/dg/speeches/detail/who-director-general-s-opening-remarks-at-the-media-briefing-on-covid-19---11-march-2020},
  [Online; accessed 30-June-2020] (2020).

\bibitem{liu2020covid}
Y.-C. Liu, R.-L. Kuo, S.-R. Shih, Covid-19: The first documented coronavirus
  pandemic in history, Biomedical journal (2020).

\bibitem{worldometer}
Worldometer, Covid-19 coronavirus pandemic,
  \url{https://www.worldometers.info/coronavirus}, [Online; accessed
  30-June-2020] (2020).

\bibitem{sulaiman2020dynamical}
A.~Sulaiman, On dynamical analysis of the data-driven sir model (covid-19
  outbreak in indonesia), medRxiv (2020).

\bibitem{zhao2020estimating}
S.~Zhao, S.~S. Musa, Q.~Lin, J.~Ran, G.~Yang, W.~Wang, Y.~Lou, L.~Yang, D.~Gao,
  D.~He, et~al., Estimating the unreported number of novel coronavirus
  (2019-ncov) cases in china in the first half of january 2020: a data-driven
  modelling analysis of the early outbreak, Journal of clinical medicine 9~(2)
  (2020) 388.

\bibitem{roda2020difficult}
W.~C. Roda, M.~B. Varughese, D.~Han, M.~Y. Li, Why is it difficult to
  accurately predict the covid-19 epidemic?, Infectious Disease Modelling
  (2020).

\bibitem{lin2020conceptual}
Q.~Lin, S.~Zhao, D.~Gao, Y.~Lou, S.~Yang, S.~S. Musa, M.~H. Wang, Y.~Cai,
  W.~Wang, L.~Yang, et~al., A conceptual model for the outbreak of coronavirus
  disease 2019 (covid-19) in wuhan, china with individual reaction and
  governmental action, International journal of infectious diseases (2020).

\bibitem{tang2020estimation}
B.~Tang, X.~Wang, Q.~Li, N.~L. Bragazzi, S.~Tang, Y.~Xiao, J.~Wu, Estimation of
  the transmission risk of the 2019-ncov and its implication for public health
  interventions, Journal of clinical medicine 9~(2) (2020) 462.

\bibitem{yang2020modified}
Z.~Yang, Z.~Zeng, K.~Wang, S.-S. Wong, W.~Liang, M.~Zanin, P.~Liu, X.~Cao,
  Z.~Gao, Z.~Mai, et~al., Modified seir and ai prediction of the epidemics
  trend of covid-19 in china under public health interventions, Journal of
  Thoracic Disease 12~(3) (2020) 165.

\bibitem{fanelli2020analysis}
D.~Fanelli, F.~Piazza, Analysis and forecast of covid-19 spreading in china,
  italy and france, Chaos, Solitons \& Fractals 134 (2020) 109761.

\bibitem{salgotra2020time}
R.~Salgotra, M.~Gandomi, A.~H. Gandomi, Time series analysis and forecast of
  the covid-19 pandemic in india using genetic programming, Chaos, Solitons \&
  Fractals (2020) 109945.

\bibitem{djilali2020coronavirus}
S.~Djilali, B.~Ghanbari, Coronavirus pandemic: A predictive analysis of the
  peak outbreak epidemic in south africa, turkey, and brazil, Chaos, Solitons
  \& Fractals (2020) 109971.

\bibitem{roosa2020real}
K.~Roosa, Y.~Lee, R.~Luo, A.~Kirpich, R.~Rothenberg, J.~Hyman, P.~Yan,
  G.~Chowell, Real-time forecasts of the covid-19 epidemic in china from
  february 5th to february 24th, 2020, Infectious Disease Modelling 5 (2020)
  256--263.

\bibitem{li2020propagation}
L.~Li, Z.~Yang, Z.~Dang, C.~Meng, J.~Huang, H.~Meng, D.~Wang, G.~Chen,
  J.~Zhang, H.~Peng, et~al., Propagation analysis and prediction of the
  covid-19, Infectious Disease Modelling 5 (2020) 282--292.

\bibitem{acuna2020modeling}
M.~A. Acu{\~n}a-Zegarra, M.~Santana-Cibrian, J.~X. Velasco-Hernandez, Modeling
  behavioral change and covid-19 containment in mexico: A trade-off between
  lockdown and compliance, Mathematical Biosciences (2020) 108370.

\bibitem{feng2020benefits}
Z.~Feng, J.~W. Glasser, A.~N. Hill, On the benefits of flattening the curve: A
  perspective, Mathematical Biosciences (2020) 108389.

\bibitem{dhanwant2020forecasting}
J.~N. Dhanwant, V.~Ramanathan, Forecasting covid 19 growth in india using
  susceptible-infected-recovered (sir) model, arXiv preprint arXiv:2004.00696
  (2020).

\bibitem{bertozzi2020challenges}
A.~L. Bertozzi, E.~Franco, G.~Mohler, M.~B. Short, D.~Sledge, The challenges of
  modeling and forecasting the spread of covid-19, arXiv preprint
  arXiv:2004.04741 (2020).

\bibitem{de2020sir}
C.~A. De~Castro, Sir model for covid-19 calibrated with existing data and
  projected for colombia, arXiv preprint arXiv:2003.11230 (2020).

\bibitem{qi2020model}
C.~Qi, D.~Karlsson, K.~Sallmen, R.~Wyss, Model studies on the covid-19 pandemic
  in sweden, arXiv preprint arXiv:2004.01575 (2020).

\bibitem{shim2020transmission}
E.~Shim, A.~Tariq, W.~Choi, Y.~Lee, G.~Chowell, Transmission potential and
  severity of covid-19 in south korea, International Journal of Infectious
  Diseases (2020).

\bibitem{benvenuto2020application}
D.~Benvenuto, M.~Giovanetti, L.~Vassallo, S.~Angeletti, M.~Ciccozzi,
  Application of the arima model on the covid-2019 epidemic dataset, Data in
  brief (2020) 105340.

\bibitem{Johns}
Johns hopkins university center for systems science and engineering,
  \url{https://github.com/CSSEGISandData/COVID-19}, accessed: 2020-07-21.

\bibitem{fattah2018forecasting}
J.~Fattah, L.~Ezzine, Z.~Aman, H.~El~Moussami, A.~Lachhab, Forecasting of
  demand using arima model, International Journal of Engineering Business
  Management 10 (2018) 1847979018808673.

\bibitem{calafiore2020modified}
G.~C. Calafiore, C.~Novara, C.~Possieri, A modified sir model for the covid-19
  contagion in italy, arXiv preprint arXiv:2003.14391 (2020).

\bibitem{kucharski2020early}
A.~J. Kucharski, T.~W. Russell, C.~Diamond, Y.~Liu, J.~Edmunds, S.~Funk, R.~M.
  Eggo, F.~Sun, M.~Jit, J.~D. Munday, et~al., Early dynamics of transmission
  and control of covid-19: a mathematical modelling study, The lancet
  infectious diseases (2020).

\bibitem{peng2020epidemic}
L.~Peng, W.~Yang, D.~Zhang, C.~Zhuge, L.~Hong, Epidemic analysis of covid-19 in
  china by dynamical modeling, arXiv preprint arXiv:2002.06563 (2020).

\bibitem{wangping2020extended}
J.~Wangping, H.~Ke, S.~Yang, C.~Wenzhe, W.~Shengshu, Y.~Shanshan, W.~Jianwei,
  K.~Fuyin, T.~Penggang, L.~Jing, et~al., Extended sir prediction of the
  epidemics trend of covid-19 in italy and compared with hunan, china,
  Frontiers in medicine 7 (2020) 169.

\bibitem{chatterjee2020healthcare}
K.~Chatterjee, K.~Chatterjee, A.~Kumar, S.~Shankar, Healthcare impact of
  covid-19 epidemic in india: A stochastic mathematical model, Medical Journal
  Armed Forces India (2020).

\bibitem{liang2020mathematical}
K.~Liang, Mathematical model of infection kinetics and its analysis for
  covid-19, sars and mers, Infection, Genetics and Evolution (2020) 104306.

\bibitem{ndairou2020mathematical}
F.~Ndairou, I.~Area, J.~J. Nieto, D.~F. Torres, Mathematical modeling of
  covid-19 transmission dynamics with a case study of wuhan, Chaos, Solitons \&
  Fractals (2020) 109846.

\bibitem{hellewell2020feasibility}
J.~Hellewell, S.~Abbott, A.~Gimma, N.~I. Bosse, C.~I. Jarvis, T.~W. Russell,
  J.~D. Munday, A.~J. Kucharski, W.~J. Edmunds, F.~Sun, et~al., Feasibility of
  controlling covid-19 outbreaks by isolation of cases and contacts, The Lancet
  Global Health (2020).

\bibitem{van2017reproduction}
P.~van~den Driessche, Reproduction numbers of infectious disease models,
  Infectious Disease Modelling 2~(3) (2017) 288--303.

\bibitem{fraser2009pandemic}
C.~Fraser, C.~A. Donnelly, S.~Cauchemez, W.~P. Hanage, M.~D. Van~Kerkhove,
  T.~D. Hollingsworth, J.~Griffin, R.~F. Baggaley, H.~E. Jenkins, E.~J. Lyons,
  et~al., Pandemic potential of a strain of influenza a (h1n1): early findings,
  science 324~(5934) (2009) 1557--1561.

\bibitem{rodpothong2012viral}
P.~Rodpothong, P.~Auewarakul, Viral evolution and transmission effectiveness,
  World Journal of Virology 1~(5) (2012) 131.

\bibitem{ma2020estimating}
J.~Ma, Estimating epidemic exponential growth rate and basic reproduction
  number, Infectious Disease Modelling 5 (2020) 129--141.

\bibitem{roberts2007model}
M.~Roberts, J.~Heesterbeek, Model-consistent estimation of the basic
  reproduction number from the incidence of an emerging infection, Journal of
  mathematical biology 55~(5-6) (2007) 803.

\bibitem{bettencourt2008real}
L.~M. Bettencourt, R.~M. Ribeiro, Real time bayesian estimation of the epidemic
  potential of emerging infectious diseases, PLoS One 3~(5) (2008) e2185.

\bibitem{forsberg2008likelihood}
L.~Forsberg~White, M.~Pagano, A likelihood-based method for real-time
  estimation of the serial interval and reproductive number of an epidemic,
  Statistics in medicine 27~(16) (2008) 2999--3016.

\bibitem{obadia2012r0}
T.~Obadia, R.~Haneef, P.-Y. Bo{\"e}lle, The r0 package: a toolbox to estimate
  reproduction numbers for epidemic outbreaks, BMC medical informatics and
  decision making 12~(1) (2012) 1--9.

\bibitem{wallinga2004different}
J.~Wallinga, P.~Teunis, Different epidemic curves for severe acute respiratory
  syndrome reveal similar impacts of control measures, American Journal of
  epidemiology 160~(6) (2004) 509--516.

\bibitem{fine2011herd}
P.~Fine, K.~Eames, D.~L. Heymann, Herd immunity: a rough guide, Clinical
  infectious diseases 52~(7) (2011) 911--916.

\bibitem{ganyani2020estimating}
T.~Ganyani, C.~Kremer, D.~Chen, A.~Torneri, C.~Faes, J.~Wallinga, N.~Hens,
  Estimating the generation interval for covid-19 based on symptom onset data,
  MedRxiv (2020).

\end{thebibliography}
